\documentclass[aps,prl,twocolumn,groupedaddress,showpacs,floatfix]{revtex4}
\def\bc{\begin{center}}
\def\ec{\end{center}}
\def\be{\begin{equation}}
\def\ee{\end{equation}}

\usepackage{longtable}
\usepackage{epsfig}

\begin{document}

\title{\bf Microscopic origin of the next generation fractional quantum Hall effect}
\author{C.-C Chang and J.K. Jain}
\address{Department of Physics, 104 Davey Laboratory, The Pennsylvania State University,
Pennsylvania 16802}
\date{\today}

\begin{abstract}

Most of the fractions observed to date belong to
the sequences $\nu=n/(2pn\pm 1)$ and $\nu=1-n/(2pn\pm 1)$, $n$ and $p$ integers, 
understood as the familiar {\em integral} quantum Hall effect of 
composite fermions.  These sequences fail to accommodate, however,
many fractions such as $\nu=4/11$ and $5/13$, 
discovered recently in ultra-high mobility samples at very low 
temperatures.  We show that these ``next 
generation" fractional quantum Hall states are 
accurately described as the {\em fractional} quantum Hall effect of 
composite fermions.
\end{abstract}

\maketitle

Electrons confined in two dimensions, when subjected to a strong magnetic 
field, form a quantum fluid that exhibits the remarkable phenomena of
integral and fractional quantum Hall effects~\cite{Klitzing,Tsui},
namely quantized Hall resistance plateaus at
$R_H=h/\nu e^2$ with integral and fractional values of $\nu$.
The integral quantum Hall effect (IQHE) is explained as a property of
uncorrelated electrons, resulting from a quantization of the kinetic 
energy of electrons into Landau levels in the presence of a magnetic 
field.  The fractional quantum Hall effect (FQHE), on the other hand, is 
a manifestation of a strongly correlated quantum fluid.
At very strong magnetic fields, electrons fall into the lowest Landau level (LL) 
and the physics is entirely governed by the repulsive Coulomb interaction.
Many essential properties of this quantum fluid can be explained by postulating that 
electrons in the lowest LL minimize their interaction energy by capturing an even number 
($2p$) of quantized vortices each to turn into composite fermions (CF's)~\cite{Jain},
which experience an effective magnetic field and form their own 
Landau-like levels, termed ``CF-quasi-Landau levels."  The number of occupied 
CF-quasi-Landau levels, $\nu^*$, is related to the filling factor of 
the lowest electron LL, $\nu$, according to the  
formula $\nu=\nu^*/(2p\nu^*\pm 1)$.  In particular, 
the IQHE of composite fermions ($\nu^*=n$) provides an explanation for  
the FQHE of electrons at $R_H=h/\nu e^2$, with $\nu=n/(2pn\pm 1)$.
(Particle hole symmetry in the lowest LL gives $1-\nu$.)

The observation of fractions such as 4/11 and 5/13~\cite{Pan,Pan2,Smet,Goldman} 
points to new physics beyond the integral quantum Hall 
effect of composite fermions.  It has been 
appreciated that the residual interaction between composite fermions can, 
in principle, cause such fractions~\cite{Jain,Jain2}, in the same way 
as the interaction between electrons produces the FQHE.  For example, consider 
composite fermions carrying two vortices ($p=1$).  If they were completely
non-interacting, only $\nu=n/(2n\pm 1)$ would be obtained.
However, there {\em is} a weak residual interaction between 
composite fermions.  If it happens to be of a form that 
produces a {\em fractional} QHE of composite fermions at 
\begin{equation}
\nu^*=1+\bar{\nu}=1+\frac{m}{2m\pm 1}\; 
\label{nu*}
\end{equation}
($m=$integer), then that would result in new {\em electron} fractions in the range 
$2/5>\nu>1/3$, given by
\begin{equation}
\nu=\frac{3m\pm 1}{8m\pm 3}\;.
\label{nu}
\end{equation}
These include the newly observed fractions~\cite{Comment}.
(Of course, many more fractions can be constructed in this manner~\cite{Jain2}.)

This qualitative picture is intuitively appealing, 
and indicates that the next generation FQHE is possible at least for some model 
interaction between composite fermions.  However,  to confirm this scenario,
it is important to carry out quantitative tests to determine 
if the FQHE of composite fermions will occur for 
the {\em actual} residual interaction between 
composite fermions, a remnant from the Coulomb interaction between electrons.
A FQHE at $\nu$ requires that the state here be 
incompressible, that is, have a uniform 
ground state with a gap to excitations.  One can ascertain 
incompressibility from either exact numerical diagonalization 
on small systems, or ``CF diagonalization"~\cite{Mandal}
(outlined below) for larger systems.  Extensive studies 
of $\nu=4/11$ as a function of the number of electrons, $N$, have found that 
the state is incompressible for $N=12$ and 20  
but compressible for $N=8$, 16 and 24~\cite{Comm1,Xie,Wojs,Rezayi,Mandal}.
While the message was mixed, it was on the whole interpreted to mean that the results  
do not support, in the thermodynamic limit, a fully spin-polarized 
FQHE at $\nu=4/11$~\cite{Mandal}.  That conclusion, however,
is incompatible with experiments~\cite{Pan,Pan2}, which show a clear evidence 
for a fully polarized FQHE at $\nu=4/11$.
We explain below the origin of the intriguing behavior for finite systems, 
why it does {\em not} rule out incompressibility in the 
thermodynamic limit (contrary to our previous assertion),  and then go on 
to write explicit wave functions to confirm, quantitatively, that the 
new fractions indeed are well described as the FQHE of composite 
fermions.

The calculations below will consider $N$ electrons on the 
surface of a sphere, moving under the influence of a radial 
magnetic field $B$ produced by a Dirac monopole of strength 
$Q$ at the center, which produces a net magnetic flux of $2Q$, in units of 
$\phi_0=hc/e$, through the surface. ($2Q$ is an integer according Dirac's 
quantization condition).  ``CF diagonalization" refers to determining 
the low-energy spectrum by numerically diagonalizing the Hamiltonian in 
the correlated CF basis:
\begin{equation}
\{{\cal P}_{LLL}\Phi_1^{2p}\Phi^{\alpha}_{Q^*}\}
\label{basis}
\end{equation}
Here $\{\Phi^\alpha_{Q^*}\}$ is an orthogonal basis of 
$N$-electron states at $Q^*=Q-p(N-1)$.
We will be interested in the filling factor 
range $2/5>\nu>1/3$, for which the CF filling at 
$\nu^*$ lies between one and two (with $p=1$).
We will include all electron basis states at $Q^*$  
which have the lowest LL completely occupied and the second 
partially occupied.  $\Phi_1$ is the wave function for a fully
occupied Landau level, and ${\cal P}_{LLL}$ denotes projection 
into the lowest Landau level.  The basis states in 
Eq.~(\ref{basis}) are in general not linearly independent.
We extract an orthogonal basis following the Gram-Schmidt
procedure and then diagonalize the Coulomb Hamiltonian to find 
eigenstates and eigenenergies.  The Hamiltonian matrix elements 
are evaluated by Metropolis Monte Carlo method.
(All energies will be quoted in units of $e^2/\epsilon l_0$, where
$\epsilon$ is the dielectric constant of the background semiconductor, 
and $l_0\equiv \sqrt{\hbar c/eB}$ is the magnetic length.)
The statistical uncertainty is determined 
by performing many ($\sim$10) Monte Carlo runs, with $0.8-1.0 \times 10^6$ 
iterations in each run.  The basis states are, by construction, in the 
lowest Landau level, so our results provide strict variational bounds on the 
ground state energy in the limit $B\rightarrow \infty$. 
The eigenstates have definite orbital angular momentum, $L$, 
with $L=0$ for uniform ground states.
The ground state obtained by CF diagonalization 
will be denoted $\Psi_{\nu}^{0}$.
Details of lowest LL projection and diagonalization can be 
found in the literature~\cite{JK,Mandal}. 
The state at filling factor $\nu$ of Eq.~(\ref{nu}) is obtained 
at flux values given by~\cite{Mandal}
\be
2Q=\frac{8m\pm 3}{3m\pm 1}N-\frac{12m\pm(m^2+3)}{3m\pm 1}
\ee
which ensures $\lim_{N\rightarrow\infty}N/2Q=\nu=(3m\pm 1)/(8m\pm 1)$.
It has been shown in the past that the CF diagonalization method 
produces essentially the same results as exact diagonalization (see, 
for example, Ref.~\onlinecite{Jainssc}).

We begin by pointing out the flaw in the reasoning of Ref.~\onlinecite{Mandal}
that led to the conclusion that the fully spin-polarized state at $\nu=4/11$ etc. 
is compressible in the thermodynamic limit.  It was implicitly assumed in Ref.~\onlinecite{Mandal}  
that if a state is incompressible in the thermodynamic limit, then 
all its finite size realizations must also be incompressible.  This criterion was 
used because there was no known exception to it for FQHE in the lowest Landau level, 
at fractions of the form $\nu=n/(2pn\pm 1)$.
However, the criterion is not universally valid, and FQHE states in 
{\em higher} LL's provide an explicit counter-example.  Consider the electron state at  
$\nu^*=1+\bar{\nu}$, with $\bar{N}$ particles forming a state with filling factor 
$\bar{\nu}$ in the second LL.  Given that the filled lowest LL is inert,
one might expect that the state in the second LL at $\bar{\nu}$ is 
quite similar to the corresponding state at filling factor $\bar{\nu}$ in the lowest LL, 
but, in reality, there are striking differences between the two~\cite{Reynolds,Haldane2},
for reasons not fully understood at present.  Consider the example of $\bar{\nu}=1/3$. 
For the 1/3 state in the lowest LL, the system is incompressible for all $N$, 
whereas for the 1/3 state in the second LL, the ground state 
is compressible ($L\neq 0$) for $\bar{N}=3$ and 5, and almost compressible
for $\bar{N}=7$.  (See Ref.~\onlinecite{Reynolds} and Table I.   
The gap for $\bar{N}=7$ is a factor of 37 smaller than the gap at $\bar{N}=6$.)
Furthermore, the ground state wave functions at 1/3 in the lowest and 
second LL's are rather different; the largest overlap between them 
is obtained for seven particles, which is only $\sim$0.6~\cite{Reynolds}.  
Because of such strong fluctuations as a function of $N$ 
it was initially thought~\cite{Haldane2} 
that exact diagonalization studies {\em rule out} FQHE at $\bar{\nu}=1/3$ in 
the second LL.  Study of bigger systems revived the possibility  
of incompressibility in the thermodynamic limit~\cite{Reynolds}, and FQHE 
at 1/3 in the second LL has been observed experimentally~\cite{Stormer},
albeit with a small gap of $\sim$100 mK.

Could something similar be happening at the newly observed fractions?
That would be quite natural from the CF perspective, which relates the new fractions 
(e.g. $\nu=4/11$) to the FQHE of composite fermions in {\em higher} CF-quasi-LL's
(e.g. $\nu^*=4/3$).  We now proceed to investigate the issue quantitatively.  
To begin with, Table I gives the gaps at several values of $\nu$ given by Eq.~\ref{nu}, 
obtained by CF diagonalization.  (No gap is given when the ground state is not 
uniform.)  The behavior is remarkably analogous to that at $\nu^*=1+\bar{\nu}$ 
(Eq.~\ref{nu*}).  For example, including the electrons in the lowest LL, 
the state at $\nu^*=4/3$ is 
compressible for $N=8$ and 16 particles ($\bar{N}=3$ and 5) and almost 
compressible for $N=24$ ($\bar{N}=7$); these match the particle numbers for 
which $\nu=4/11$ has been found to be compressible.
The states at 5/13 and 7/19 are similar to those at 5/3 and 7/5.  
The analogy between $\nu$ and $\nu^*$ 
strongly suggests that, in spite of finite size fluctuations, the 
state at $\nu$ is incompressible in the thermodynamic limit. 
It would be desirable to study systems at 
$\nu=(3m\pm 1)/(8m\pm 3)$ larger than those reported here and in Ref.~\onlinecite{Mandal}, 
but that is not possible with the present day computers.

\begin{table}
\begin{tabular}{cccccccc}
\hline\hline
$\nu$ & $\nu^*$ & $2Q$ & $N$ & $\bar{N}$  & gap ($\nu$) & gap ($\nu^*$) \\
\hline
               & & 18 & 8 & 3 &  -       & - \\
$\frac{4}{11}$ & $\frac 4 3$ & 29 & 12 & 4    & 0.010(1) & 0.035 \\
               & & 40 & 16 & 5   & -       & - \\
               & & 51 & 20 & 6   & 0.006(2) & 0.024 \\
               & & 62 & 24 & 7   & -         & 0.00064       \\
\hline
$\frac{5}{13}$ & $\frac 5 3$ & 33 & 14 & 6   & 0.003(1) & 0.035 \\
               & & 46 & 19 & 8    & -       & - \\
\hline
               & & 20 & 9 & 4  & -       & - \\
$\frac{7}{19}$ & $\frac 7 5$ & 39 & 16 & 6   & 0.006(2) & 0.016 \\
               & & 58 & 23 & 8  & -       & 0.0018 \\
\hline\hline
\end{tabular}
\caption{The gaps at $\nu$ and $\nu^*$, determined from CF and exact diagonalization,
respectively, at several filling factors.  Only the gaps of incompressible 
states are shown.  $N$ is the total number of particles and $\bar{N}$ is 
the number of particles in the second LL for the state at $\nu^*$.
The gaps are quoted in units of $e^2/\epsilon l_0$.
The statistical uncertainty from Monte Carlo is shown in parentheses.
}
\end{table}

We now concentrate on those particle numbers for which the states
are incompressible, which we believe contain the physics of incompressibility 
in the thermodynamic limit.
A secure understanding of the origin of a FQHE state rests 
on identifying an accurate wave function that reveals its microscopic physics.  
Wave functions for the new FQHE states can be constructed based on the above 
physical picture following the standard procedure, which allows for a microscopic 
test of the scenario.  For the  ground state at 
$\nu=(3m\pm 1)/(8m\pm 3)$ the trial wave function is given by 
\be
\Psi_{\nu}^{tr} = {\cal P}_{LLL} \Phi_1^2 \Phi_{\nu^*}
\ee
where $\Phi_{\nu^*}$ is the $L=0$ Coulomb ground state at $\nu^*=1+m/(2m\pm 1)$,
obtained by exact diagonalization.  
Because multiplication by $\Phi_1^2$ attaches two vortices to each electron 
to convert it into a composite fermion, the wave function $\Psi_{\nu}^{tr}$
is interpreted as the FQHE of composite fermions at $\nu^*=1+m/(2m\pm 1)$.
(Although amenable to an intuitive 
interpretation through composite fermions, the actual wave 
function is extremely intricate.)  As another reference point, we will also 
present results for the trial wave function
\be
\Psi^{'tr}_{\nu} = {\cal P}_{LLL} \Phi_1^2 \Phi'_{\nu^*}
\ee
where $\Phi'_{\nu^*}$ is obtained by placing in the second LL the 
Coulomb ground state at $\bar{\nu}=m/(2m\pm 1)$ of the {\em lowest} Landau level.
$\Psi^{tr}_{\nu}$ is derived from the $m/(2m\pm 1)$ state in the second LL, 
whereas $\Psi^{'tr}_{\nu}$ is analogous to the $m/(2m\pm 1)$ state in the lowest LL.

Because $\Psi_{\nu}^0$ are very accurate~\cite{Jainssc}, 
the overlaps and energies given in Table II establish that 
$\Psi^{tr}_{\nu}$ are also very accurate.  (The overlap of 
0.86 is significantly large for a system with $N=20$ particles.)
A direct comparison with {\em exact} 
results is possible for the 12 particle state at $\nu=4/11$.  For this system, 
the energies from the CF theory, $E^0=-0.44105(9)$ and $E^{tr}=-0.44088(4)$, 
deviate from the exact energy, $-$$0.441214$~\cite{Rezayi} by 0.04\% and 0.08\%.
The level of agreement is highly significant for a system with 12 particles,
and similar to that for the accepted 
trial wave functions for the ordinary FQHE at $\nu=n/(2pn\pm 1)$.
It gives an unambiguous indication, at a microscopic level, of a direct connection between 
the physics of the FQHE at $\nu$ and $\nu^*$.

\begin{table}
\begin{tabular}{ccccccc}
\hline\hline
$\nu$ &  $N$ &  ${\cal O}$ & ${\cal O}'$ & $E^0$ & $E^{tr}$ & $E^{'tr}$  \\ \hline
$\frac{4}{11}$  &  12 & 0.993(2)    & 0.51(1) & -0.44105(9) & -0.44088(4)& -0.43670(4) \\
                &  20 & 0.86(1)  & 0.278(8) & -0.43027(5)& -0.42975(5) & -0.42705(6) \\ \hline
$\frac{5}{13}$ &  14  & 0.973(1)     & 0.365(3) &-0.44400(9)& -0.44374(9) & -0.43951(3) \\ \hline
$\frac{7}{19}$ &  16  & 0.990(4)     & 0.009(2) & -0.43808(4)& -0.43806(4)& -0.43283(5)
 \\ \hline
\hline
\end{tabular}
\caption{Comparison of trial wave functions 
$\Psi^{tr}_{\nu}$ and $\Psi^{'tr}_{\nu}$  with  
$\Psi_{\nu}^0$ for several incompressible states 
at three filling factors.  (See text for definitions.)  The overlaps 
are defined as ${\cal O}=\langle \Psi_\nu^0|\Psi_\nu^{tr} \rangle/
\sqrt{\langle \Psi_\nu^0|\Psi_\nu^{0} \rangle\langle \Psi_\nu^{tr}|\Psi_\nu^{tr} \rangle}$ 
and ${\cal O}' =\langle \Psi_\nu^0|\Psi_\nu^{'tr} \rangle /\sqrt{\langle \Psi_\nu^0|\Psi_\nu^0
\rangle \langle \Psi_\nu^{'tr}|\Psi_\nu^{'tr} \rangle}$.  
$E^0$, $E^{tr}$ and $E^{'tr}$ are the Coulomb energies per particle for 
$\Psi_{\nu}^0$, $\Psi^{tr}_{\nu}$ and $\Psi^{'tr}_{\nu}$,
respectively.  $E^0$ were reported in Ref.~\protect\onlinecite{Mandal}. 
}
\end{table}

Thus, the analogy between the FQHE 
in the lowest LL at $\nu$ (Eq.~\ref{nu}) and the FQHE in the second LL 
at $\nu^*=1+\bar{\nu}$ (Eq.~\ref{nu*}) not only explains the qualitative behavior 
as a function of $N$, but also produces accurate wave functions 
for the {\em incompressible} ground states at $\nu$.  These facts taken together 
give us confidence that the new fractions are a manifestation of  
the FQHE of composite fermions.

Even though the finite system incompressible states help us confirm the physics
of the new fractions, an  accurate determination of the excitation gaps is not possible 
because we do not have enough points for a reliable extrapolation to the thermodynamic limit. 
The gap in Table I for the largest available incompressible system at a given filling
can be taken as a very crude estimate for the thermodynamic gap.
The gaps for the new FQHE states are more than an order of magnitude smaller than the 
gaps at $\nu=1/3$ and 2/5 in the lowest LL, 
explaining why the new FQHE states are much more fragile, more readily destroyed by 
disorder or thermal fluctuations, than the ordinary FQHE states at 1/3 and 2/5, 
in spite of their close proximity.
The smallness of the gap is illustrative of the fact that the 
residual interaction between composite fermions is much weaker than the interaction 
between electrons.  (All gaps being compared are theoretical gaps, without 
including the effects of finite thickness or disorder.)

There has been much recent theoretical work on the new FQHE states.
Pairing of composite fermions has been advanced~\cite{Quinn} as an  
alternative possible mechanism for the next generation fractions, 
and another theoretical paper~\cite{Goerbig} has studied the FQHE 
of composite fermions using a Hamiltonian approach~\cite{Murthy}.
However, the quantitative accuracy of the methods used in these works has not been established at 
a level required for the issue of stability of the delicate new FQHE states, and 
in particular, neither of these approaches constructs an explicit 
wave function which can be directly compared with the exact ground state wave 
function known for small systems.
An effective field theory approach~\cite{Fradkin} is also unable to address 
the stability of a FQHE state, although it may 
illuminate certain properties thereof assuming it exists.  
%They work with $\bar{N}$ composite
%fermions in the second CF-quasi-LL; they assume that the two-body interaction 
%is dominant, which is estimated by considering the system with only two composite 
%fermions in the second CF-quasi-LL~\cite{Sitko,Lee} (which has filling factor 
%$\nu\approx 1/3$).  
%Surely, the actual form of the two-body interaction will change as  
%the filling factor moves away from $\nu=1/3$, and the terms involving 
%three and more composite fermions will also become relevant.
%Our method does not make any such assumptions regarding the form of the 
%interaction between composite fermions, deals with the full $N$ particle 
%problem, and does not find any evidence for pairing. 
%The physics described in our work is not 
%Our explicit wave functions, demonstrably accurate, do not show any evidence for 
%based on any pairing.
%Exact diagonalization for this model problem 
%shows that the system is incompressible at $2Q=(11/4)N-5$, which 
%is interpreted in terms of pairing of composite fermions.  

We discuss briefly certain approximations made in our work.  (i) Our 
main assumption is the neglect of mixing between the CF-quasi-LL's.
%The basis states at $Q^*$, from which the correlated CF basis is 
%obtained, does not allow for any LL mixing.  
Our preliminary studies, which relax 
the assumption by enlarging the basis (by allowing for LL mixing at $Q^*$),
find that the corrections 
are very small and do not change the qualitative results.  
Indeed, the fact that $\Psi^0_\nu$ is very close to the exact ground 
state is indicative of the insignificance of CF-quasi-LL mixing. 
(ii) We are also neglecting, throughout, a mixing between {\em electronic} 
Landau levels at $Q$, but that is presumably negligible at the highest 
magnetic fields where the next generation FQHE has been 
observed~\cite{Pan,Pan2}. 
(iii) We have also studied the correction due to a finite transverse 
thickness of the electron system, which modifies the form of the effective 
two-dimensional interaction between electrons; it lowers all energies 
but does not alter significantly either the qualitative nature 
of the ground state or the form of the ground state wave function. 
(iv) The Zeeman energy has been assumed to be frozen.  
Spin related physics can also generate 
new fractions, associated with partially spin polarized states,
which would be observable at relatively low magnetic 
fields.  These are actually more straightforward 
to understand theoretically than the fully spin polarized states~\cite{Chang}, 
because they are analogous to FQHE 
in the {\em lowest} LL.  For example, $\nu=4/11$ FQHE would be related to 
FQHE at $\nu^*= 1 (\uparrow) + \bar{\nu} (\downarrow)$, where 
now the spin up states of the lowest LL are fully occupied and spin down 
states of the lowest LL have a fractional filling of $\bar{\nu}$.  
The states observed in Refs.~\onlinecite{Pan,Pan2} 
are insensitive to changes in the Zeeman energy and survive to  
very high magnetic fields, indicating that they are fully spin polarized. 
An early hint of fractions outside the sequences $n/(2pn\pm 1)$ 
in very low density samples~\cite{Goldman} may involve partial spin reversal.

An extension of the above analogy between FQHE at $\nu$ and $\nu^*$ 
has implications for future fractions.
There is good evidence~\cite{Gervais} that the Coulomb interaction does not stabilize 
FQHE of electrons at $\nu=\bar{n}+m/(2m\pm 1)$ for $\bar{n}>1$, but 
fractions like $\bar{\nu}=1/5$ and $\bar{\nu}=4/5$ may occur in the third LL.
Assuming similar behavior for composite fermions, this would rule out 
fully spin polarized FQHE at electron fractions of the form 
$[(2\bar{n}+1)m\pm \bar{n}]/[4(\bar{n}+1)m\pm(2\bar{n}+1)]$
with $\bar{n}>1$, but leave open the possibility of FQHE at apparently 
more complicated fractions like $\nu=11/27$ and $14/33$ in the filling
factor range $2/5<\nu<3/7$.  Charge density waves of various types are also 
predicted to occur for certain filling factors in higher quasi-LL's of 
composite fermions~\cite{Lee}.

%In summary, we have shown that the next generation FQHE  
%can be understood quantitatively as the fractional quantum Hall effect of 
%composite fermions, in which the composite fermions in the second CF-quasi-LL 
%transform into higher order composite fermions to 
%form a FQHE state.  The puzzling 
%compressibility for certain finite size systems is explained 
%as being due to strong finite size effects, similar to those found 
%previously for the ordinary FQHE in higher electronic LL's.  This physical 
%understanding of the new fractions is verified quantitatively through 
%construction of explicit wave functions.

We thank M.R. Peterson for useful discussions
and E.H. Rezayi for sharing with us his exact diagonalization results.
Partial support of this research by the National Science Foundation under grant
no. DMR-0240458 is acknowledged.


\begin{references}
\bibitem{Comm1} Exact spectra for eight 
particles in the filling factor range between 1/3 and 2/5 
were first presented in Ref.~\onlinecite{Xie}.  Ref.~\onlinecite{Wojs}
noted that the eight particle state at $\nu=4/11$ was not incompressible.  
Recently, Rezayi~\cite{Rezayi} has 
obtained exact results for 12 particles at 4/11 (which is incompressible).  
Systems with up to 24 particles were studied in Ref.~\onlinecite{Mandal} by CF 
diagonalization.
\end{references}
\end{document}